\documentclass[slac_one]{revtex4}
\usepackage{graphicx}
\usepackage{fancyhdr}
\newcommand{\mev}{\text{MeV}}
\newcommand{\gev}{\text{GeV}}
\newcommand{\tev}{\text{TeV}}

\newcommand{\m}{\text{m}}

\newcommand{\s}{\text{s}}

\newcommand{\eqref}[1]{Eq.~(\ref{#1})}

\newcommand{\figref}[1]{Fig.~\ref{fig:#1}}

\newcommand{\kkh}{H^{\pm\, 1}}

\newcommand{\mH}{m_{\kkh}}

\begin{document}

\title{{\small{LEPTON - PHOTON 2007}}\\ 
\vspace{12pt}
Minimal Universal Extra Dimensions} 

\author{Jose~A.~R.~Cembranos}
\affiliation{Department of Physics and Astronomy, University of
California, Irvine, CA 92697, USA}

\author{Jonathan L.~Feng}
\affiliation{Department of Physics and Astronomy, University of
California, Irvine, CA 92697, USA}

\author{Louis E.~Strigari}
\affiliation{Department of Physics and
Astronomy, University of California, Irvine, CA 92697, USA}

\begin{abstract}
Highly degenerate spectra associated with universal extra dimensions (UED)
provide an interesting phenomenology not only from the point of view of 
cosmology and astrophysics, but also for colliders. We study these exotic 
signals for the simplest case, called minimal UED, where it is natural to 
find slow charged particles, displaced vertices, tracks with non-vanishing
impact parameters, track kinks, and even vanishing charged tracks.
\end{abstract}

\maketitle


\thispagestyle{fancy}

\section*{Introduction}

Minimal universal extra dimensions (mUED) is defined as the minimum consistent
extension of the standard model (SM) with additional spatial dimensions. All
the particles propagate in one flat, compact extra dimension 
compactified on an $S^1/Z_2$ orbifold of size
$10^{-18}~\m$ or smaller~\cite{KK}~\cite{Appelquist:2000nn}. 

The properties of the model depend on only one new parameter at tree level, 
the size of the extra dimension, $R$. This model is non renormalizable, and 
one needs to introduce a cut-off, $\Lambda$, that we will fix to the standard value 
$\Lambda R=20$ \cite{Cheng:2002iz}. The other free parameter of this model is the 
last free parameter of the SM: the Higgs mass $m_h$. In this model, the particle 
spectrum is naturally highly degenerate. Inside the allowed region of the parameter
space of mEUD by standard cosmology, the lightest Kaluza-Klein (KK) particle (LKP) 
and the next-to-lightest KK particle (NLKP) are the KK gauge boson, $B^1$,  and the graviton, 
$G^1$, with maximal mass splittings of   $\sim 1 \gev$ \cite{Cembranos:2006gt}. 
The LKP is completely stable by KK parity and the lifetime of the NLKP is given by the formula:
\begin{eqnarray}
\tau& \simeq &
\frac{3\pi}{b} 
\frac{M_P^2}{(\Delta m)^3}\simeq
\frac{3.57 \times 10^{22}~\s}{b}
\left[\frac{\mev}{\Delta m} \right]^3\,,
\label{eq:gravidecay}
\end{eqnarray}
where  $M_P$ is the reduced Planck scale $M_P\equiv (8\pi G_N)^{-1/2}\simeq 2.4 \times 10^{18}$ GeV.
The $b$ factor is of order one and depends on which particle is the lightest one:
$b= 10 \cos^2\theta_W /3\simeq 2.54$ for the $B^1 \to G^1 \gamma$ decay, and
$b= 2 \cos^2\theta_W \simeq 1.52$ for the $G^1 \to B^1 \gamma$ decay \cite{Feng:2003xh} 
($\theta_W$ is the weak mixing angle, $\cos^2\theta_W \simeq 0.76$). 

The lifetime of these NLKPs is longer than the age of the Universe for 
compactification scales $795$ GeV $<R^{-1}< 820$ GeV. This region is 
particularly important for $G^1$ LKPs ($R^{-1}< 809$ GeV). In this region of 
parameter space, the thermal $B^1$ abundance has the preferred value 
observed for the
non-baryonic dark matter (for Higgs masses $180$ GeV $<m_h< 215$ GeV), and 
it is not excluded by present observations (see Ref.~\cite{Kakizaki:2006dz,Cembranos:2006gt} 
and Fig. \ref{fig:summary}).

\section*{Constraints}

The size of the extra dimension has an upper bound from 
precision electroweak measurements: $R^{-1} \agt
250~\gev$~\cite{Appelquist:2000nn,Appelquist:2002wb}, with other low
energy constraints similar or
weaker~\cite{Agashe:2001xt,Appelquist:2001jz}.  Particle physics alone
does not place any lower bound on $R$, but the thermal relic
density of LKPs grows with $R^{-1}$, and LKPs would overclose the
universe for $R^{-1} > 1.5~\tev$~\cite{Servant:2002aq,%
Kakizaki:2005en,Kakizaki:2005uy,Burnell:2005hm,Kong:2005hn,%
Kakizaki:2006dz}, providing strong motivation for considering
weak-scale KK particles.  Direct contrains on the standard model 
Higgs boson mass apply to mUED requiring $m_h >
114.4~\gev$ at 95\% CL~\cite{Barate:2003sz}.  In contrast, however,
the indirect bounds on $m_h$ are significantly weakened relative to
the standard model, requiring only $m_h < 900~\gev$ for $R^{-1} =
250~\gev$ and $m_h < 300~\gev$ for $R^{-1} = 1~\tev$ at 90\%
CL~\cite{Appelquist:2002wb}. However, if $m_h > 245~\gev$, the lightest
KK particle is the charged higgs boson, whose relic abundace rules out the model unless the reheat temperature is very low.

\section*{Lightest particles}

Early studies of UED focused on the line in model parameter space
defined by $m_h = 120~\gev$~\cite{Cheng:2002ab} and neglected the
existence of the KK graviton $G^1$~\cite{Servant:2002aq,Cheng:2002ej}.
Given these assumptions, for $R^{-1} \agt 250~\gev$, the LKP is the
hypercharge gauge boson $B^1$, and these studies therefore focused on
missing energy signals at colliders and weakly-interacting massive
particle (WIMP) dark matter for cosmology.  These predictions are
similar to those from supersymmetry with $R$-parity conservation. UED
with KK-parity and supersymmetry with $R$-parity predict different
collider event rates for similar spectra, and the different spins of
partner particles may be distinguished through, for example, indirect
dark matter detection in positrons~\cite{Cheng:2002ej}.  Nevertheless,
the difficulty of distinguishing UED and supersymmetry has attracted
much attention and been a fertile testing ground for future
experiments, especially the Large Hadron Collider
(LHC)~\cite{Macesanu:2005jx}.

\begin{figure}
\resizebox{3.56in}{!}{
\includegraphics{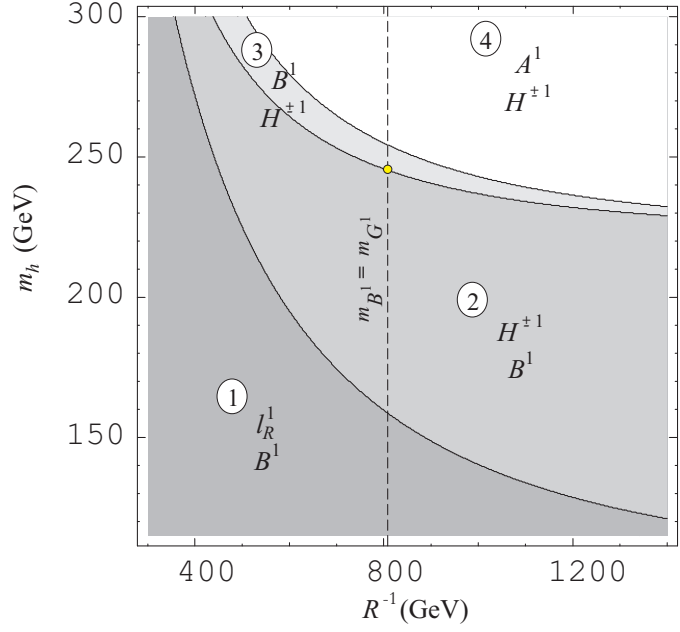}
} 
\caption{The complete collider phase diagram of mUED in the $(R^{-1},
m_h)$ plane, where $R$ is the compactification radius, and $m_h$ is
the Higgs boson mass.  The KK graviton $G^1$ has been excluded from
consideration, and the standard model (NLKP, LKP) pairs in each phase
are as indicated.  We have set $\Lambda R = 20$.  For reference, the
line on which $m_{G^1} = m_{B^1}$ is also plotted.  The ``triple
point,'' where $m_{B^1} = m_{G^1} = \mH$, is at $(R^{-1}, m_h) \approx
(810~\gev, 245~\gev)$ \cite{Cembranos:2006gt}.
\label{fig:phases} }
\end{figure}

More recent studies have shown that framework of UED
is far richer than indicated above.  First, it was noted that the KK
graviton $G^1$ necessarily exists in any UED model and may be the LKP,
leading not to WIMP dark matter, but to superWIMP dark matter, with a
completely different set of cosmological and astroparticle
signatures~\cite{Feng:2003xh,Feng:2003uy,Feng:2003nr}.  Second,
studies have now emphasized that, by relaxing the constraint $m_h =
120~\gev$ and considering higher values, KK Higgs bosons may become
lighter than the $B^1$.  That both of these possibilities may be
realized in a general UED model is, perhaps, not surprising.
Remarkably, however, all of these complexities arise even in the
extremely constrained framework of mUED.  Any one of the $G^1$, $B^1$,
and the charged Higgs boson $\kkh$ may be the LKP, leading to many
different ``phases'' of parameter space with qualitatively distinct
signatures.  The ``triple point,'' where $m_{G^1} = m_{B^1} =
m_{\kkh}$, lies in the heart of parameter space at $(R^{-1}, m_h)
\approx (810~\gev, 245~\gev)$, leading to many interesting features.

\section*{Exotic collider phenomenology associated with long-lived particles}

\begin{figure}
\resizebox{3.56in}{!}{
\includegraphics{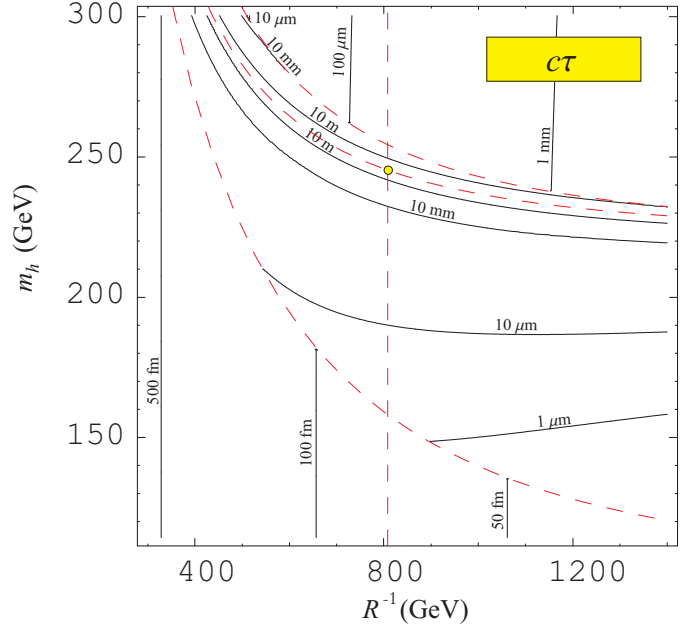}
} 
\caption{Standard model NLKP decay lengths in the full phase space of
mUED. We have set $\Lambda R = 20$ and neglected the KK graviton
$G^1$. The dashed curves are the boundaries of phases shown in
\figref{phases} \cite{Cembranos:2006gt}.
\label{fig:lifetimes} }
\end{figure}

The degeneracies throughout the mUED phase diagram are typically of $1\%$ or $0.1\%$.  
These degeneracies suppress NLKP decay widths, such that NLKPs produced 
in colliders may decay at points macroscopically separated from the interaction point. 
In fact, if we wet aside the KK graviton $G^1$ 
and cosmological considerations, mUED supports 4 distinct standard model (NLKP, LKP) 
combinations, or phases, as it is shown in Fig. \ref{fig:phases}.
The NLKP decay lengths throughout parameter space are
given in \figref{lifetimes}.  

In addition, strongly-interacting KK particles may be produced with large rates at
the LHC since the KK spectrum is highly degenerate.  Long-lived NLKP tracks will 
therefore presumably be most easily identified in the cascade decays of KK quarks and gluons.
Such events will be characterized by many jets and missing transverse
energy, which will satisfy trigger criteria, and the jets will fix the
interaction point.  The possible signals are \cite{Cembranos:2006gt}:
\begin{itemize}
\item Phase 1: Prompt decays $l_R^1 \to B^1 l_R$, where $l = e, \mu,
  \tau$, the mass splitting between KK states is $\Delta m \sim {\cal
  O}(\gev)$, and the final state lepton is consequently very soft.
\item Phase 2: Decays $\kkh \to B^1 f \bar{f}'$, where $f \bar{f}' =
  e^+ \nu_e, \mu^+ \nu_{\mu}, u \bar{d}, \tau^+ \nu_{\tau}, c
  \bar{s}$, where the decay length is $c \tau \agt 100~\text{nm}$ (for
  $R^{-1} \alt 1400~\gev$) and may be effectively infinite for
  collider phenomenology.  Again $\Delta m \sim {\cal O}(\gev)$, and
  the final state fermions are very soft.  Depending on the
  observability of the final state fermions, the exotic signatures
  could include non-prompt decays producing displaced vertices, tracks
  with non-vanishing impact parameters, track kinks, or even
  disappearing charged ($\kkh$) tracks that mysteriously vanish after
  passing through only part of the detector.  In the parameter region
  where the $\kkh$ is effectively stable, it may be produced at low
  velocities, resulting in time-of-flight anomalies and
  highly-ionizing tracks.
\item Phase 3: Decays $B^1 \to \kkh f \bar{f}'$, where the $f
  \bar{f}'$ pairs are as in Phase 2, with decay length typically
  satisfying $c \tau \agt 10~\text{mm}$ (except in a tiny region, in
  which could be even shorter than $10 \mu\m$), and again $\Delta m
  \sim {\cal O}(\gev)$, and the final state standard model fermions
  are very soft. The possible signatures are as above, with the
  exception that, since the NLKP is neutral and the LKP is charged in
  this case, NLKP events could instead be seen as charged ($\kkh$)
  tracks that mysteriously {\em appear} somewhere in the detector.
\item Phase 4: Decays $A^1 \to \kkh f \bar{f}$, where $f = e, \mu,
  \tau, u, d, s, c$, and the decay length is constrained to the
  relatively narrow range $10~\mu\m \alt c \tau \alt 1~\text{mm}$.
  The signatures are similar to Phase 3.
\end{itemize}

\section*{Conclusions}

Taking into account these collider signatures, it is interesting to
study the KK graviton, $G^1$, the nature of the dark matter (DM) in the
model, and the consequences of late decays in cosmology or astrophysics. 
It has been shown that long-lived particles
~\cite{Feng:2003xh, SWIMPs} have associated a large number of possible 
astrophysical signatures, such as modifications of light element abundances 
~\cite{BBN,boundstates},  anomalies in the cosmic microwave background 
\cite{Ichiki:2004vi}, deviations from the cold DM power 
spectrum~\cite{Cembranos:2005us,Kaplinghat:2005sy,Sigurdson:2003vy}, modifications
in the reionization history \cite{Chen:2003gz} or anomalies in cosmic ray
spectra \cite{Cembranos:2007fj}. The analysis of all these signals, complements other 
astrophysical searches \cite{others} and related collider experiments~\cite{CollSW,Coll}. 

\begin{figure}
\resizebox{4.56in}{!}{
\includegraphics{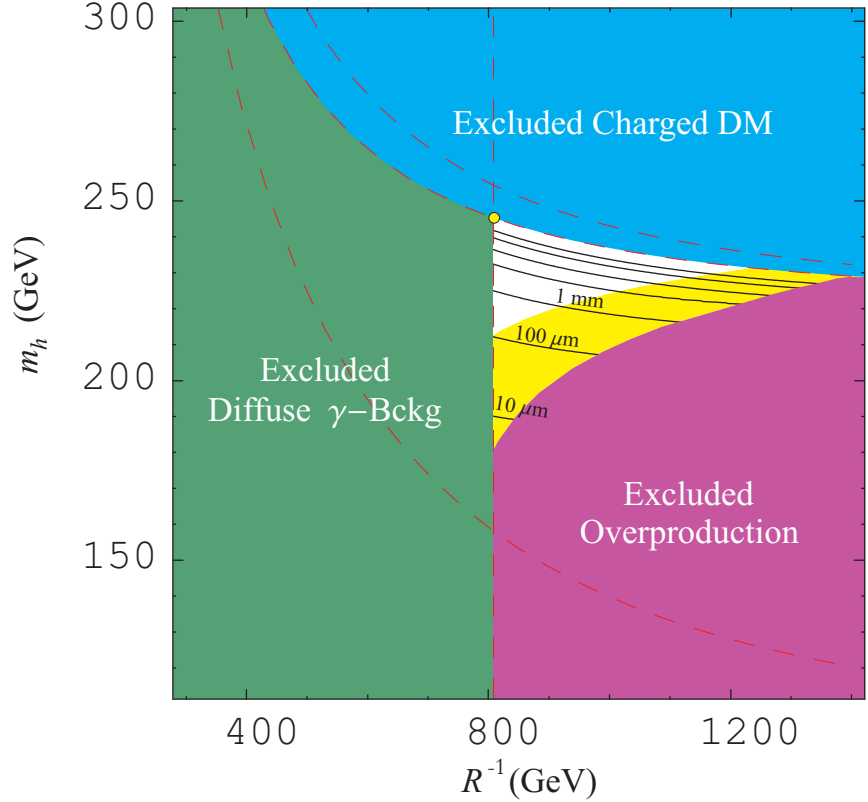}
} 
\caption{The cosmologically preferred region of the complete phase
diagram of mUED.  The $G^1$ has been included, and the dark shaded
regions are excluded by the cosmological constraints on stable charged
relics, the diffuse photon flux, and WIMP overproduction, as
indicated.  In the preferred region, the light shaded region is from
Ref.~\cite{Kakizaki:2006dz} and shows where the $B^1$ thermal relic
density is within $2\sigma$ of the WMAP central value for non-baryonic
dark matter.  Contours of constant decay length $c \tau = 10~\mu\m,
100~\mu\m, 1~\text{mm}, \ldots, 1~\m, 10~\m$ are also plotted (only
the lowest few are labeled) \cite{Cembranos:2006gt}.
\label{fig:summary} }
\end{figure}
  
One of these signals may be the first evidence of this scenario. At the same time,
they introduce important constraints to the model as we have briefly commented.
For a low enough reheating temperature after inflation, all 4 phases,
even Phases 3 and 4 with charged LKPs, are viable. On the contrary, 
much of the phase diagram is excluded if one assumes
a standard cosmology with reheat temperature above $R^{-1}/25$.  The
final results are given in \figref{summary}. Phases 3 and 4 are
excluded by bounds on stable charged particles, Phases 1 and 2 with
$R^{-1} < 810~\gev$ are excluded by bounds from the observed diffuse
MeV photon flux, and Phases 1 and 2 with high $R^{-1}$ are excluded
because WIMPs are overproduced through thermal freeze-out.  

The resulting cosmologically preferred region is bounded on all sides 
as it can be seen in \figref{summary}.  In this region the Higgs boson 
mass lies in the range $180~\gev \alt m_h \alt 245~\gev$,  which 
is allowed and implies the ``golden'' 4 lepton signatures for Higgs bosons at the
LHC.  The compactification radius satisfies $810~\gev \alt R^{-1} \alt
1400~\gev$ and  The LKP mass is approximately in this range, and the
heaviest $n=1$ KK particle is never heavier than 320 GeV, which means that
KK particles will be copiously produced at the LHC. On the contrary, none of
these new particles would be produced directly at the International
Linear Collider operating at center of mass energies below 1.5 TeV.
The decay $\kkh \to B^1 f \bar{f}'$ has associated decay lengths 
$c\tau \agt 4~\mu\m$, without upper bound.  Generically, then, 
long-lived tracks are expected at the LHC.
This phenomenology is very distinctive of mUED in relation to other theories
beyond the standard model such as supersymmetry, where these signatures require
fine-tunning among the different parameters of the model.

\begin{acknowledgments}
The work of JARC and JLF is supported in part by NSF CAREER grant No.~PHY--0239817,
NASA Grant No.~NNG05GG44G, and the Alfred P.~Sloan Foundation. The work of
JARC is also supported by the FPA 2005-02327 project (DGICYT, Spain).
LES and JARC are supported in part by a Gary McCue Postdoctoral Fellowship
through the Center for Cosmology at UC Irvine.
\end{acknowledgments}


\end{document}